\documentclass[journal,11pt,draftclsnofoot,onecolumn]{IEEEtran}
\ifCLASSINFOpdf
\else
\fi
\usepackage{graphicx}
\usepackage{subfigure}
\usepackage{amsmath}
\usepackage{amsmath,amssymb,amsthm}

\begin{document}

\title{Stochastic Process-based Method for Degree-Degree Correlation of Evolving Networks}

\author{\IEEEauthorblockN{Yue~Xiao\IEEEauthorrefmark{2}, Xiaojun~Zhang\thanks{* Xiaojun~Zhang is the corresponding author.}\IEEEauthorrefmark{3}}\\ 
	\IEEEauthorblockA{ School of Mathematical Science, University of Electronic Science and Technology of China\\ Chengdu, Sichuan 611731, China\\
 }}
\maketitle

\begin{abstract}
Existing studies on the degree correlation of evolving networks typically rely on differential equations and statistical analysis, resulting in only approximate solutions due to inherent randomness. To address this limitation, we propose an improved Markov chain method for modeling degree correlation in evolving networks. By redesigning the network evolution rules to reflect actual network dynamics more accurately, we achieve a topological structure that closely matches real-world network evolution. Our method models the degree correlation evolution process for both directed and undirected networks and provides theoretical results that are verified through simulations. This work offers the first theoretical solution for the steady-state degree correlation in evolving network models and is applicable to more complex evolution mechanisms and networks with directional attributes. Additionally, it supports the study of dynamic characteristic control based on network structure at any given time, offering a new tool for researchers in the field.
\end{abstract}


\IEEEpeerreviewmaketitle

\textbf{\section{Introduction}}
The degree correlation of complex networks is an important statistic for describing the topological structure of the network. Unlike the degree distribution, which measures the number of edges connected to a single node, degree correlation describes the tendency of nodes to connect to other nodes based on their degree values. This statistic reflects the average connection pattern within the network and is usually quantified using the Pearson correlation coefficient of the degrees at the endpoints of each edge. When high-degree nodes tend to connect to other high-degree nodes, the network exhibits assortative mixing, or positive degree correlation. In contrast, when high-degree nodes are more likely to connect to low-degree nodes, the network exhibits disassortative mixing, or negative degree correlation. Existing research has extensively explored these patterns. For example, Newman (2002) showed that social networks generally exhibit assortative mixing, while technical and biological networks tend to have disassortative mixing\cite{newman2002assortative}.\\
\indent Network degree correlation has a significant impact on network dynamics, with different connection patterns profoundly affecting network functionality. In the study of network percolation\cite{noh2007percolation}, synchronization control\cite{di2007effects,sendina2015effects,yu2021network,ming2017effect}, and robustness research\cite{bonneau2020statistical}, researchers often introduce an edge reconnection mechanism into network models with a given degree distribution. This approach allows for numerical and theoretical analysis of network dynamics before and after changes, enabling the evaluation of the impact of degree correlation on these dynamics. For instance, in percolation research, degree correlation influences the critical threshold at which a network transitions from a fragmented to a connected state. In synchronization control, the degree correlation affects how quickly and effectively networked oscillators can synchronize. Robustness studies have shown that assortative networks (where similar nodes are connected) tend to be more resilient to random failures but more vulnerable to targeted attacks, whereas disassortative networks (where dissimilar nodes are connected) exhibit different resilience characteristics. These dynamic features have broader applications in real-world networks. For example, in social networks, assortativity enhances information dissemination and social influence. In biological and technological networks, heteroassortativity (disassortativity) helps to understand network robustness and efficiency. Additionally, degree correlation plays a critical role in thermal transport within complex networks\cite{xiong2018effect}, strategic interactions in network games\cite{lotfi2022effect}, and the dynamics of epidemic transmission\cite{morita2023representation}. Understanding degree correlation provides insights into the structure and behavior of networks that go beyond what is captured by degree distribution alone. This deeper understanding helps identify potential vulnerabilities and design more resilient and efficient networks.\\
\indent Traditionally, addressing degree correlation in networks has focused on static spatial network models. These models analyze the structural characteristics of networks at a fixed point in time, particularly focusing on the impact of spatial distribution and physical distance between nodes\cite{bertotti2019configuration,williams2014degree, bertotti2019configuration,mondragon2020estimating,yao2017average,fujiki2018general,antonioni2012degree,van2013degree, jing2015disassortative}. It is worth noting that some studies have shown that using the Pearson correlation coefficient to characterize network degree correlation can be insufficiently rigorous. For instance, in fixed-scale scale-free networks, the lower bound of degree correlation is not $-1$\cite{yang2017lower}. Furthermore, incorporating degree correlation is essential for addressing fundamental inverse problems in network science. When the network structure is unknown, obtaining a network with complex structural characteristics using only the degree distribution is challenging. Therefore, it is crucial to incorporate degree correlation because it provides local connection information that is vital for capturing characteristics such as community size boundaries, shortest path length, and clustering coefficient\cite{luo2022random,jones2022clarifying}.\\
\indent However, the structure of real networks often changes over time as nodes and edges are added or deleted. Although discussions on the degree correlation of evolving networks have been ongoing, the characterization indicators and analysis methods vary. For growing networks with uniform node deletion, the degree distribution and degree correlation change over time and influence each other. Some studies use continuous methods to determine the average nearest neighbor degree under certain degree distribution conditions\cite{garcia2008degree}. Similarly, when the network evolution mechanism includes preferential node deletion, the discussion of degree correlation becomes more complex. In such cases, some studies use rate methods to explore the degree distribution when degrees are uncorrelated\cite{juher2011uncorrelatedness}. In scale-free network models, the rate method has been used to theoretically solve the joint probability $P(k_1,k_2)$ as a degree correlation indicator. However, this approach assumes that nodes are uncorrelated and the degree distribution of each node's neighbors is independent of the node's characteristics\cite{fotouhi2013degree}. Recognizing that real-world evolutionary mechanisms can produce different degree correlations alongside power-law distributions, some studies use the master equation method to deeply study the time-dependent relationship of $P(k_1,k_2)$ and verify through simulations\cite{stamos2013evolution}. For models like the Barabási-Albert (BA) model and the ternary closure model based on preferential attachment, the average nearest neighbor degree (ANND) is used to quantify degree correlation. Studies have derived the distribution of ANND at each iteration from the mean field equation\cite{mironov2021degree}. In addition to conventional methods, some researchers have employed matrix eigenvalue decomposition to study the interaction between degree correlation and epidemic behavior, aiding in the analysis of social network dynamics\cite{morita2023representation}. In addition, some studies have used Markov chains to characterize the change process of the average degree of neighboring nodes. This method has been successfully applied to real network data, providing insights into the variability of local characteristics in complex networks\cite{sidorov2023measuring}.

\indent Comprehensive comparison shows that researchers have tried to use different methods to measure the degree correlation of evolving networks, but it is still difficult to clearly grasp the node connection situation of the network based on the approximate results obtained. Because compared with static networks, the evolutionary network process introduces random factors, which makes the network topology structure at any time have multiple possibilities and the proportion of networks with certain special structures is larger. Static networks with fixed network topology are like samples randomly drawn at any time during the evolution of dynamic networks, and the discussion of the degree correlation of evolving networks is more complicated. Especially when the evolutionary mechanism includes node deletion, the degree values of its neighboring nodes will decrease, and the number of these neighboring nodes is related to the degree correlation of the deleted node; if node preference deletion is performed, the network degree distribution at the next moment is more related to the degree distribution of the deleted node in the network at this moment. Choosing appropriate degree correlation measurement indicators for evolving networks and considering how to model the theoretical results of degree correlation are areas where research in this field can be improved. Therefore, this paper takes the simplest pure growth network as an example to study the degree correlation under the two attributes of directed and undirected. This work can provide a theoretical basis for the study of degree correlation under more complex evolutionary mechanisms in the future. The contents of each chapter are as follows: Section 2 gives different definitions of degree correlation. Section 3 introduces the pure growth network model studied in this paper. Section 4 proposes the evolution process of directed networks under the Markov chain method and the corresponding degree correlation transfer equation. Section 5 briefly gives the degree correlation results of undirected networks based on the discussion of degree correlation of directed networks. Section 6 is a summary of the work in this paper and prospects for future research.\\

\section{Preliminaries}\label{sec.2}
In network science, correlation usually refers to the relationship between the properties of nodes or edges within a network. The Pearson correlation coefficient \(r\) is proposed to measure the tendency of nodes to connect to other nodes with similar or different degrees; the average neighbor degree function \( k_{nn}(k) \) describes the correlation between the degree of a node and the average degree of its neighbors; aim to represent the probability of a connection between nodes of a certain degree, the degree-degree correlation \( P(k, k') \) also be proposed in researches. We give introduction of these indicators specifically as below.\\

\textbf{$1.$ Correlation coefficient (\(r\))}\\
Pearson correlation coefficient of connecting node degrees is defined as:
\begin{equation}
 r = \frac{E[k_i k_j] - E[k_i]E[k_j]}{\sqrt{(E[k_i^2] - (E[k_i])^2)(E[k_j^2] - (E[k_j])^2)}} 
\end{equation}
 where \( k_i \) and \( k_j \) are the degrees of the nodes at both ends of the edge, and \( E[\cdot] \) represents the expected value. When $r\geq 0$ the corresponding network is homogeneous mixing, that is, nodes with high degrees tend to connect to other nodes with high degrees. When $r\leq 0$ the network is heterogeneous mixing, means nodes with high degrees tend to connect to nodes with low degrees.\\
 
\textbf{$2.$ Average neighbor degree function \( k_{nn}(k) \)}\\
The average degree of neighbors of a node with degree \( k \) is defined as:
\begin{equation}
 k_{nn}(k) = \frac{1}{N_k} \sum_{i|k_i=k} \frac{1}{k_i} \sum_{j \in \mathcal{N}(i)} k_j ,
\end{equation}  
where \( \mathcal{N}(i) \) represents the neighbors of node \( i \), and \( N_k \) is the number of nodes with degree \( k \).\\

\textbf{$3.$ Degree-Degree Correlation \( P(k, k') \)}\\
The probability that a randomly selected edge connects a node of degree \( k \) to a node of degree \( k' \) is defined as:
\begin{equation}
P(k, k') = \frac{L_{kk'}}{L}  , 
\end{equation}
where \( L_{kk'} \) is the number of edges connecting nodes with degree \( k' \) to nodes with degree \( k \), and \( L \) is the total number of edges in the network.\\
\indent Thus, we can conclude that correlation coefficient focuses specifically on the degrees of nodes and their connections and provides a single value summarizing the overall tendency for nodes to connect to similar or different degree nodes. The average neighbor degree correlation examines the relationship between a node’s degree and the average degree of its neighbors and provides a function describing how neighbor degrees vary with node degree. The degree correlation matrix provides a more granular view, detailing the connection probabilities between specific degree pairs.

\section{The Model}

A pure growth network is a primary evolutionary network, and many existing studies on its statistical characteristics focus on exponential degree distribution. For this type of network, at any time $t$, a new node is added to the network with $m$ edges attached. Specifically, $m$ edges are evenly connected to $m$ existing nodes in the network above; the new network obtained is the network at the next time $t+1$. In order to distinguish the above actual evolution rules from the improvements proposed later in this article, we give the evolution steps as follows:\\
\textbf{(1) Initial State:} $G(0)=G_{n0}(0)$, an initial network that is a complete network with $n0$ nodes;\\
\textbf{(2) Evolving Rules:} At time $t(t\geq 1)$, add a new node to $G_{t-1}$ and connect it with the old nodes in a given way; \\
\textbf{(3) Next Station:} $G(t)$, a new network at $t$.

The following is a schematic diagram of the actual evolution of the network from time $t-1$ to $t$.
\begin{figure}[!t]
	\centering
	\includegraphics[width=10cm]{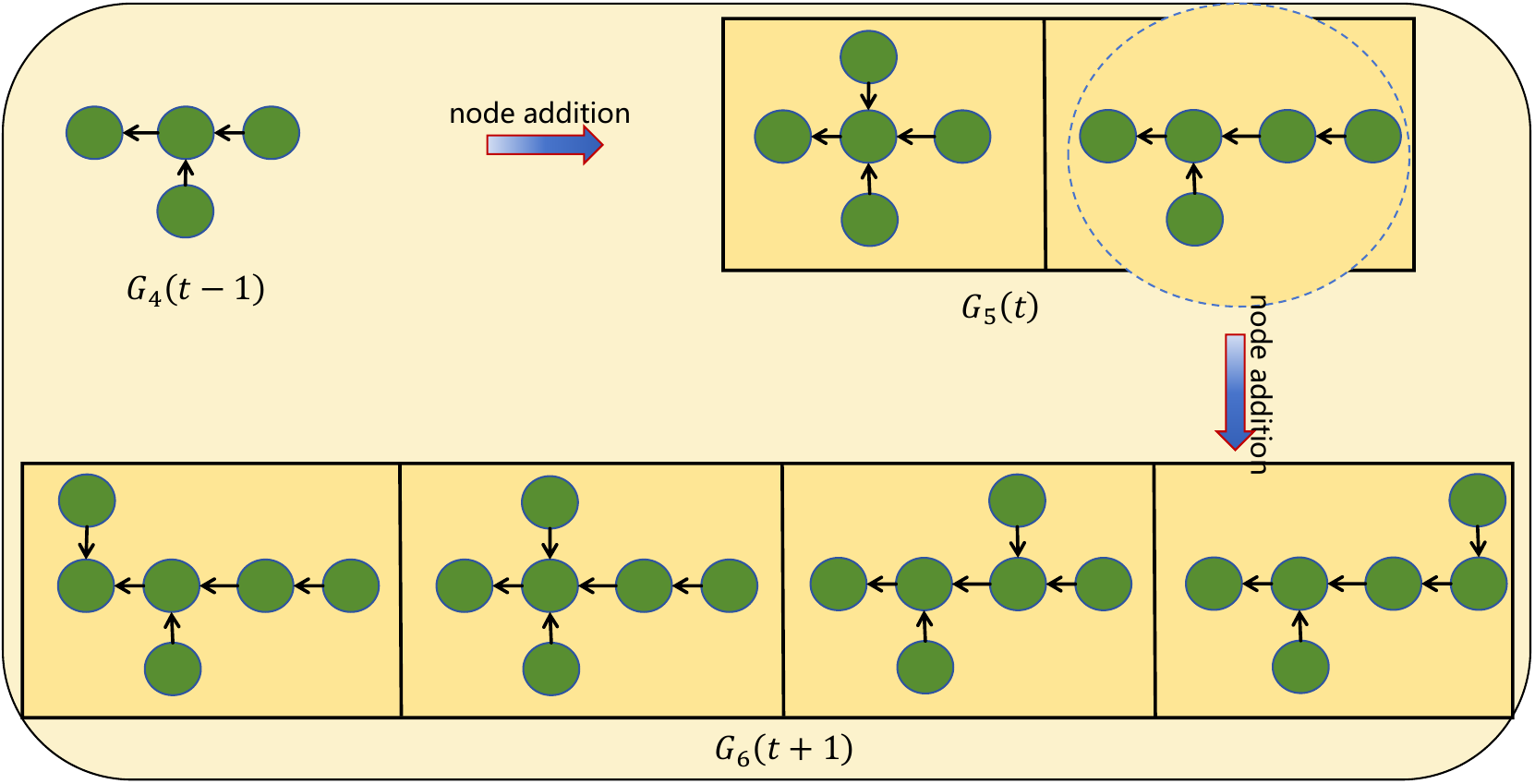}
	\caption{the actual evolution of the network.}
	\label{fig:tail}
\end{figure}\\

\section{Stochastic Process-based Approach}

In a pure growth network, with the addition of new nodes and edges, the sample space continues to change, whether from the perspective of the number of nodes or the number of edges. In order to discuss degree correlation in a sample space, we designed a transition rule based on a stochastic process(SPR) with reference to the traditional rule(TR) for evolution process to ensure the topological structure and statistical characteristics of the network. We introduce an infinite sample space $\Omega$, and we give it a partion $\Omega=\cup_{t=0}^{\infty} G(t)$. Further, we have $G(t)=\cup_{i=m}^{\infty}G_i(t)$ for each $G(t), t\geq 0$. Based on these settings, we can illustrate the evolving process of networks by the transition of edges. \\
\indent Specifically, for $m=1$ we give the corresponding approaches as follows: \\

\textbf{(1) Initial State:} $G(0)$, all the networks in $\Omega$ at $t=0$ are the complete graph with $n0$ nodes;\\
\textbf{(2) Evolving Rules:} At time $t(t\geq 1)$, for all $G_i(t-1)(i\geq m)$ in $\Omega$ \\
\indent \textbf{step1:} randomly Select $i+1$ networks from $G_i(t-1)$, then select any one of these $i+1$ networks to disassemble. Get $i$ isolated edges placed in the set $A_i (t-1)$, the remaining $i$ networks are placed in the set $B_i(t-1)$; \\

\indent \textbf{step2:} repeat the operation in step1 for other networks in $G_i(t-1)$;\\
\indent \textbf{step3:} randomly select an edge from $A_i(t-1)$, and randomly select a network from $B_i(t-1)$, and connect any endpoint of the edge to a randomly selected node in the network. Then place the obtained new network in the set $G_{i+1}(t)$;\\
\indent \textbf{step4:} repeat step 3 for other elements in $A_i(t-1)$ and $B_i(t-1)$, and place the resulting new network in $G_{i+1}(t)$.\\
\textbf{(3) Next Station:} $G(t)$, a set of new networks that $G(t)=\cup_{i=1}^{\infty}G_i(t)$ holds.\\
\indent Our above transfer rule based on Markov chain is presented in the figure below:
\begin{figure}[!t]
	\centering
	\includegraphics[width=10cm]{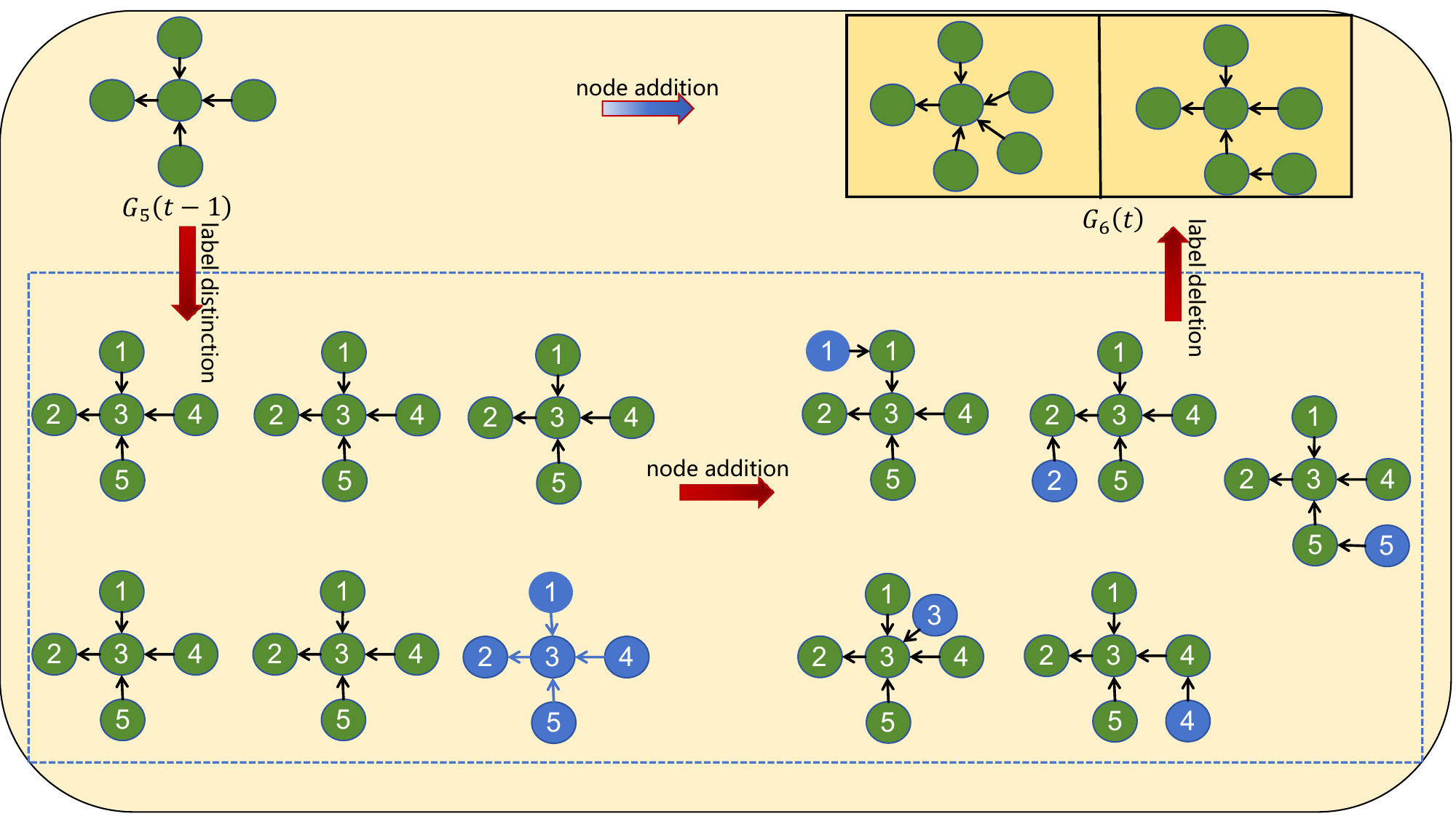}
	\caption{the evolution of the network in SPR.}
	\label{fig:tail}
\end{figure}\\

\subsection{the One-Step Transition Probability}

\indent In the network $G_{L-m}(t)$ with $L-m$ edges at time $t$, for any edge $l$, set its state to $[L-m,(k_i,k_j)]$, that is, the state of $l$ The degree value of the starting point is $k_i$, and the degree value of the endpoint is $k_j$. Therefore, in the undirected network $G_{L-m}(t)$, the state setting satisfies symmetry, and $P_{[L-m,(k_i,k_j)]}(t)=P_{[L-m,(k_i,k_j)]}(t)+P_{[L-m,(k_j,k_i)]}(t)$ represents the probability of the edge having two endpoints with degree of $k_i$ and $k_j$ at $t$; moreover, for these edges with the same degree value $k_i=k_j$ at both endpoints, the default state is $[L-m,(k_i,k_i)]$ directly. \\
\indent For the network evolution mechanism in this article, a new node is added to the network at each moment and is randomly connected to $m$ existing nodes in the network. The state of any edge can determine the change in degree correlation during the evolution of the network. Specifically, under the SPR framework, $l$ will have two states during the evolution process from $t$ to $t+1$: non-isolated and isolated. We give the corresponding transition probabilities as follows. \\

\subsubsection{$l$ is non-isolated}
\begin{equation}
	\begin{aligned}
&P_{[L-m,(k_1,k_2)][L,(k_1,k_2)]}=\frac{L-m}{L}\times\frac{{N-2\choose m}}{{N\choose m}}; \\   \quad\quad\quad  &P_{[L-m,(k_1,k_2)][L,(k_1+1,k_2)]}=\frac{L-m}{L}\times\frac{{N-2\choose m-1}}{{N\choose m}};\\
&P_{[L-m,(k_1,k_2)][L,(k_1,k_2+1)]}=\frac{L-m}{L}\times\frac{{N-2\choose m-1}}{{N\choose m}};\\
&P_{[L-m,(k_1,k_2)][L,(k_1+1,k_2+1)]}=\frac{L-m}{L}\times\frac{{N-2\choose m-2}}{{N\choose m}}
\end{aligned}
\end{equation} \\     
\subsubsection{$l$ is isolated}
\begin{equation}
	P_{[L-m,(k_1,k_2)][L,(m,k^{'})]}=\frac{m}{L}\times\frac{N_{k^{'}-1}}{N};  \\
\end{equation} \\ 

\indent Where $N$ represents the number of nodes in network $G_{L-m}(t)$ at $t$, $N_{k^{'}-1}$ represents the number of nodes that have $k^{'}-1$ neighbors in $G_{L-m}(t)$. With $N_{k_i}=(L-m)\times\frac{\sum_{k_j}P_{[L-m,(k_i,k_j)]}(t)+P_{[L-m,(k_i,k_i)]}(t)}{k_i}$, so we have $N=(L-m)\times\sum_{k_i}\frac{\sum_{k_j}P_{[L-m,(k_i,k_j)]}(t)+P_{[L-m,(k_i,k_i)]}(t)}{k_i}$. It is clear that the transition of degree-degree correlation is an inhomogeneous process, depending on the global information of network at $t$.\\
\indent It is worth noting that there is $P_{[L-m,(k_1,k_2)][L,(k_1,k_2+1)]}=\frac{L-m}{L}\times\frac{{N-2\choose m-1}}{{N\choose m}}$ when $k_1\leq k_2$ at $t$.

\subsection{state transition equation of directed networks}
In the undirected network we focused here, the degree correlation matrix at any time is symmetric. In order not to confuse, only the elements in the upper triangular matrix of the degree correlation at time $t+1$ are considered: $P_{[L,( k_1,k_2)]}(t+1)$ for $k_1=k_2$. The state transition equations are given as below.

$(1)$ For $k_1=m, k_2=m$:
\begin{equation}
\begin{aligned}
P_{[L,(m,m)]}(t+1)&=P_{[L-m,(m,m)]}(t)\times P_{[L-m,(m,m)][L,(m,m)]}(t)\\
&=\frac{L-m}{L}\frac{{N-2\choose m}}{{N\choose m}}P_{[L-m,(m,m)]}(t)
\end{aligned}
\end{equation}
$(2)$ For $k_1=m,k_2>k_1$:
\begin{equation}
\begin{aligned}
P_{[L,(m,k_2)]}(t+1)&=P_{[L-m,(m,k_2-1)]}(t)\times P_{[L-m,(m,k_2-1)][m,k_2]}(t)\\
&\indent+P_{[L-m,(m,k_2)]}(t)\times P_{[L-m,(m,k_2)][L,(m,k_2)]}(t)\\
&\indent+\sum_{k_i,k_j}P_{[L-m,(k_i,k_j)]}(t)\times P_{[L-m,(k_i,k_j)][L,(m,k_2)]}(t)\\
&=\frac{L-m}{L}\frac{{N-2\choose m-1}}{{N\choose m}}P_{[L-m,(m,k_2-1)]}(t)+\frac{L-m}{L}\frac{{N-2\choose m}}{{N\choose m}}P_{[L-m,(m,k_2)]}(t)\\
&\indent+\frac{m}{L}\frac{N_{k_2-1}}{N}\sum_{k_i,k_j}P_{[L-m,(k_i,k_j)]}(t)
\end{aligned}
\end{equation}
$(3)$ For $k_1>k_2,k_2=m$:
\begin{equation}
\begin{aligned}
P_{[L,(k_1,m)]}(t+1)&=P_{[L-m,(k_1-1,m)]}(t)\times P_{[L-m,(k_1-1,m)][k_1,m]}(t)\\
&\indent+P_{[L-m,(k_1,m)]}(t)\times P_{[L-m,(k_1,m)][L,(k_1,m)]}(t)\\
&=\frac{L-m}{L}\frac{{N-2\choose m-1}}{{N\choose m}}P_{[L-m,(k_1-1,m)]}(t)+\frac{L-m}{L}\frac{{N-2\choose m}}{{N\choose m}}P_{[L-m,(k_1,m)]}(t)\\
\end{aligned}
\end{equation}
$(4)$ For $k_1,k_2>m$:
\begin{equation}
\begin{aligned}
P_{[L,(k_1,k_2)]}(t+1)&=P_{[L-m,(k_1-1,k_2)]}(t)\times P_{[L-m,(k_1-1,k_2)][L,(k_1,k_2)]}(t)\\
&\indent +P_{[L-m,(k_1,k_2-1)]}(t)\times P_{[L-m,(k_1,k_2-1)][L,(k_1,k_2)]}(t)\\
&\indent +P_{[L-m,(k_1-1,k_2-1)]}(t)\times P_{[L-m,(k_1-1,k_2-1)][L,(k_1,k_2)]}(t)\\
&\indent +P_{[L-m,(k_1,k_2)]}(t)\times P_{[L-m,(k_1,k_2)][L,(k_1,k_2)]}(t)\\
&=\frac{L-m}{L}\frac{{N-2\choose m-1}}{{N\choose m}}P_{[L-m,(k_1-1,k_2)]}(t)+\frac{L-m}{L}\frac{{N-2\choose m-1}}{{N\choose m}}P_{[L-m,(k_1,k_2-1)]}(t)\\
&\indent + \frac{L-m}{L}\frac{{N-2\choose m-2}}{{N\choose m}}P_{[L-m,(k_1-1,k_2-1)]}(t)+\frac{L-m}{L}\frac{{N-2\choose m}}{{N\choose m}}P_{[L-m,(k_1,k_2)]}(t)
\end{aligned}
\end{equation}

Finally, the recursive relationship of the stationary correlation of directed networks is as follows
\begin{equation}
P_{(k_1,k_2)}= \left\{
\begin{aligned}
&0\quad\quad (k_1=k_2=m)\\
&\frac{m}{2m+1}P_{(m,k_2-1)}+\frac{1}{2m+1}P_{(k_2-1)} \quad\quad (k_1=m,k_2>k_1)\\
&\frac{m}{2m+1}P_{(k_1-1,m)}\quad\quad (k_1>k_2,k_2=m)\\
&\frac{m}{2m+1}[P_{(k_1-1,k_2)}+P_{(k_1,k_2-1)}] \quad\quad (k_1,k_2>m).
\end{aligned}
\right.
\end{equation}
Where $P(k)=\frac{1}{m+1}(\frac{m}{m+1})^{k-m} (k=m,m+1,m+2,...)$ is the probability distribution of the corresponding exponential tail network.

We introduce a unary generating function $G(x)=\sum_kP(k)x^k,G(1)=1$, and gradually solve the analytical expression of degree correlation.\\
\indent First, multiply the elements of the $m$th row by $x^m$, $x^{m+1}$, $x^{m+2}$, $x^{m+3}$... respectively.
\begin{equation}
\left\{
\begin{aligned}
&P(m,m)x^m=0x^m\\
&P(m,m+1)x^{m+1}=\frac{m}{2m+1}P(m,m)x^{m+1}+\frac{1}{2m+1}P(m)x^{m+1}\\
&P(m,m+2)x^{m+2}=\frac{m}{2m+1}P(m,m+1)x^{m+2}+\frac{1}{2m+1}P(m+1)x^{m+2}\\
&P(m,m+3)x^{m+3}=\frac{m}{2m+1}P(m,m+2)x^{m+3}+\frac{1}{2m+1}P(m+2)x^{m+3}\\
&P(m,m+4)x^{m+4}=\frac{m}{2m+1}P(m,m+3)x^{m+4}+\frac{1}{2m+1}P(m+3)x^{m+4}\\
...
\end{aligned}
\right.
\end{equation}
Sum all the above equations and denote $G_{m}(x)=\sum_{k_j}P(m,k_j)x^{k_j}$ we have
\begin{equation}
G_{m}(x)=\frac{1}{m+1}\frac{x}{2m+1-mx}[x^m+\frac{m}{m+1}x^{m+1}+(\frac{m}{m+1})^2x^{m+2}+(\frac{m}{m+1})^3x^{m+3}+...].
\end{equation}
Then, we can obtain 
\begin{equation}
\begin{aligned}
G_{m}(x)=\frac{1}{m+1}\sum_{k=m+1}^{\infty}{\sum_{i=1}^{k-m}\frac{m^{k-m-1}}{(2m+1)^i(m+1)^{k-m-i}}}x^k.
\end{aligned}
\end{equation}
Thus, the degree correlation $P(m,k_2)$ in $mth$ row can be determined:\\
\begin{equation}
\left\{
\begin{aligned}
&P(m,m)=0\\
&P(m,m+1)=\frac{1}{m+1}\frac{1}{2m+1}\\
&P(m,m+2)=\frac{1}{m+1}[\frac{m}{(2m+1)(m+1)}+\frac{m}{(2m+1)^2}]\\
&P(m,m+3)=\frac{1}{m+1}[\frac{m^2}{(2m+1)(m+1)^2}+\frac{m^2}{(2m+1)^2(m+1)}+\frac{m^2}{(2m+1)^3}]\\
&...\\
&P(m,k_2)=\frac{1}{m+1}\sum_{i=1}^{k_2-m}\frac{m^{k_2-m-1}}{(2m+1)^i(m+1)^{k_2-m-i}}.\\
\end{aligned}
\right.
\end{equation}

In the same way, by performing the above operations on the elements of the $(m+1)th$, $(m+2)th$, $(m+3)th$ ... rows, we finally get the relationship between the generating functions $G_r(x)$ of each row as follows: 
\begin{equation}
\left\{
\begin{aligned}
&G_{m}(x)=\frac{1}{m+1}\sum_{k=m+1}^{\infty}{\sum_{i=1}^{k-m}\frac{m^{k-m-1}}{(2m+1)^i(m+1)^{k-m-i}}}x^k\\
&G_{m+1}(x)=\frac{m}{2m+1-mx}G_m(x)\\
&G_{m+2}(x)=\frac{m^2}{(2m+1-mx)^2}G_m(x)\\
&...\\
&G_{r}(x)=\frac{m^{r-m}}{(2m+1-mx)^{r-m}}G_m(x),
\end{aligned}
\right.
\end{equation}
where $r$ is the row number.The coefficient of each term in function is the degree correlation between different states.

\subsection{numerical simulation}
In order to verify the rationality and correctness of the above modeling and reasoning, we used the Monte Carlo method to compare the recursive results of degree correlation. Specifically, we first calculated the average degree correlation of multiple exponential evolution networks obtained through simulation. As a result, the numerical calculation results of the degree correlation under the theoretical situation are obtained based on the recursion relationship, and the absolute difference between the two is found. When the difference is small enough for each value of $k_1$, $k_2$, then the theoretical derivation of this article is considered correct, and the performance in the coordinate diagram is that the two degree correlation grids almost overlap.

\begin{table}                                                                                
\centering                                                                                   
\begin{tabular}{|c|c|c|c|c|c|c|}                                                               
\hline      
$(k_1,k_2)$ & 1 & 2 & 3 & 4 & 5 & 6\\
\hline 
1 & 0 & 3.48e-04 & 1.18e-04 & 1.12e-05 & 1.21e-04 & 3.41e-04 \\           
\hline                                                                                       
2 & 0 & 2.31e-04 & 9.17e-05 & 1.11e-04 & 5.33e-04 & 8.89e-05 \\
\hline                                                                                       
3 & 0 & 5.36e-05 & 1.77e-04 & 1.24e-04 & 4.50e-05 & 1.12e-04 \\ 
\hline                                                                                       
4 & 0 & 1.07e-04 & 1.70e-05 & 2.64e-05 & 1.63e-04 & 2.16e-05 \\ 
\hline                                                                                       
5 & 0 & 3.28e-05 & 2.16e-05 & 7.10e-05 & 6.30e-05 & 7.51e-05 \\ 
\hline                                                                                       
6 & 0 & 1.51e-05 & 2.71e-05 & 3.50e-05 & 6.23e-05 & 7.15e-06 \\  
\hline                                                                                       
\end{tabular}                                                                                
\caption{Numerical calculation error for m=1.}              
\label{table:MyTableLabel}                                
\end{table}

\begin{table}                                                                                
\centering                                                                                   
\begin{tabular}{|c|c|c|c|c|c|c|}                                                               
\hline      
$(k_1,k_2)$ & 2 & 3 & 4 & 5 & 6  & 7\\
\hline 
 
2 & 0 & 1.11e-04 & 6.76e-05 & 4.51e-05 & 1.66e-04 & 6.13e-05 \\ 
\hline  
                                                                                 
3 & 0 & 3.67e-05 & 7.25e-06 & 4.24e-06 & 2.04e-04 & 1.40e-05 \\ 
\hline                                                                                       
4 & 0 & 1.05e-04 & 1.69e-04 & 8.46e-05 & 1.89e-04 & 2.45e-05 \\ 
\hline                                                                                       
5 & 0 & 1.20e-05 & 7.57e-05 & 1.27e-04 & 1.13e-04 & 6.69e-05 \\ 
\hline                                                                                       
6 & 0 & 4.59e-06 & 4.86e-05 & 7.05e-05 & 9.00e-05 & 1.54e-05 \\  
\hline  

7 & 0 & 7.94e-06 & 2.17e-05 & 3.79e-05 & 3.34e-05 & 3.96e-05 \\           
\hline          
\end{tabular}                                                                                
\caption{Numerical calculation error for m=2.}              
\label{table:MyTableLabel}                                
\end{table} 

\begin{table}                                                                                
\centering                                                                                   
\begin{tabular}{|c|c|c|c|c|c|c|}                                                               
\hline      
$(k_1,k_2)$ & 3 & 4 & 5 & 6  & 7 & 8\\
\hline 
3 & 0 & 1.40e-05 & 1.85e-04 & 3.26e-04 & 5.58e-05 & 1.07e-04 \\ 
\hline                                                                                       
4 & 0 & 3.15e-05 & 4.11e-05 & 3.40e-05 & 1.74e-05 & 1.05e-04 \\ 
\hline                                                                                       
5 & 0 & 2.42e-05 & 1.84e-04 & 4.22e-05 & 1.29e-04 & 8.00e-05 \\ 
\hline                                                                                       
6 & 0 & 2.73e-05 & 1.70e-05 & 4.98e-05 & 3.52e-05 & 4.37e-06 \\  
\hline  

7 & 0 & 7.09e-06 & 5.40e-05 & 2.08e-05 & 6.78e-06 & 5.75e-05 \\           
\hline       

8 & 0 & 1.31e-05 & 1.34e-06 & 4.54e-05 & 4.00e-08 & 2.94e-05 \\ 
\hline  
\end{tabular}                                                                                
\caption{Numerical calculation error for m=3.}              
\label{table:MyTableLabel}                                
\end{table} 

\begin{table}                                                                                
\centering                                                                                   
\begin{tabular}{|c|c|c|c|c|c|c|}                                                               
\hline      
$(k_1,k_2)$ & 4 & 5 & 6  & 7 & 8  & 9\\
\hline                                                                                      
4 & 0 & 3.44e-05 & 3.08e-05 & 6.33e-05 & 1.06e-04 & 1.02e-04 \\ 
\hline                                                                                       
5 & 0 & 3.17e-05 & 8.80e-06 & 1.75e-05 & 8.87e-05 & 2.74e-05 \\ 
\hline                                                                                       
6 & 0 & 3.44e-05 & 5.69e-05 & 1.62e-05 & 1.35e-04 & 9.92e-05 \\  
\hline  

7 & 0 & 1.04e-05 & 2.97e-05 & 6.90e-05 & 9.47e-05 & 3.25e-05 \\           
\hline       

8 & 0 & 9.34e-06 & 3.23e-05 & 8.77e-06 & 5.58e-05 & 9.43e-05 \\ 
\hline  

9 & 0 & 2.26e-06 & 3.01e-05 & 2.92e-05 & 7.84e-06 & 2.63e-05 \\ 
\hline  
\end{tabular}                                                                                
\caption{Numerical calculation error for m=4.}              
\label{table:MyTableLabel}                                
\end{table}

\begin{figure}[htbp]
	\centering      
 \begin{minipage}{0.49\linewidth}
 \centering
 \includegraphics[width=0.9\linewidth]{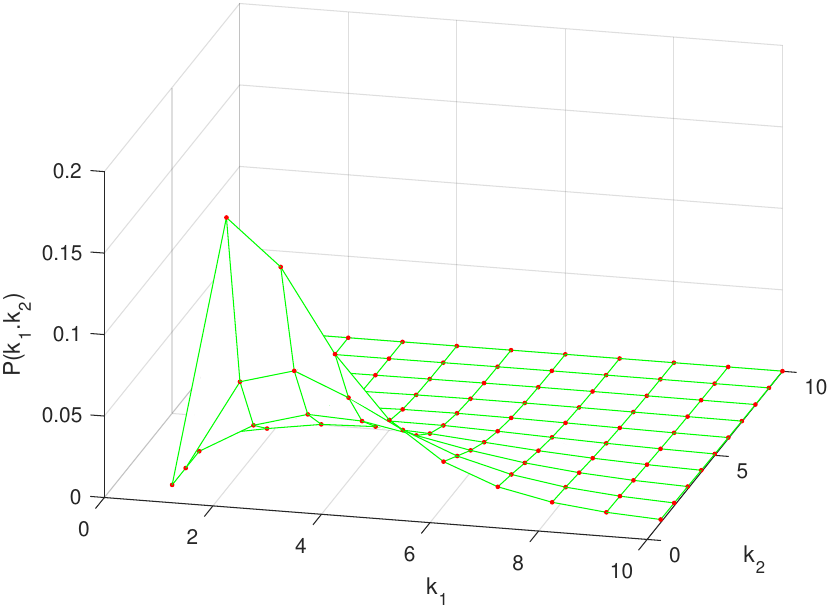}
 \end{minipage}
\begin{minipage}{0.49\linewidth}
 \centering
 \includegraphics[width=0.9\linewidth]{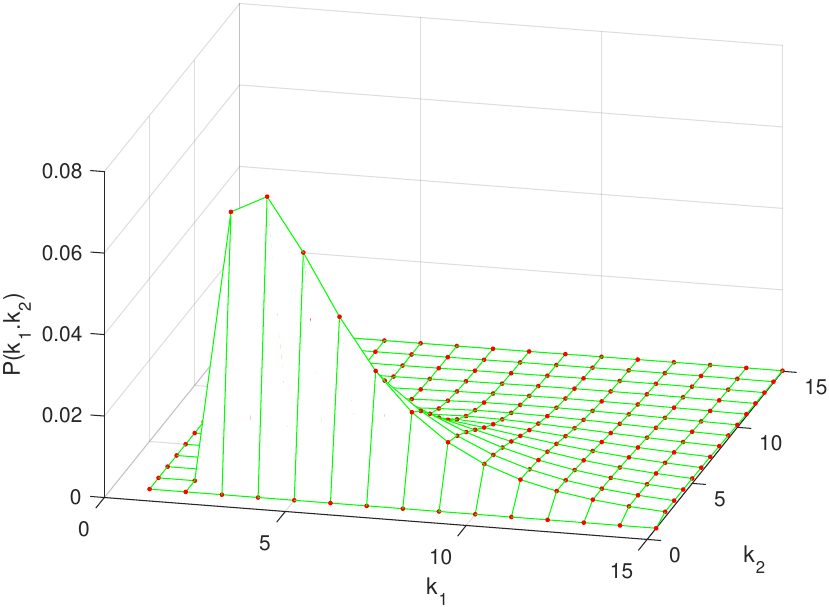}
\end{minipage}
\qquad
 \begin{minipage}{0.49\linewidth}
 \centering
 \includegraphics[width=0.9\linewidth]{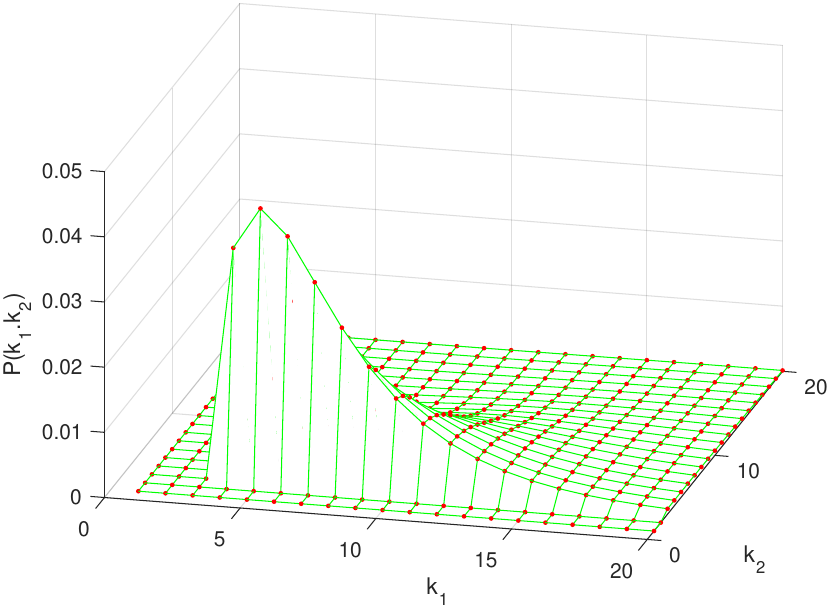}
\end{minipage}
 \begin{minipage}{0.49\linewidth}
 \centering
 \includegraphics[width=0.9\linewidth]{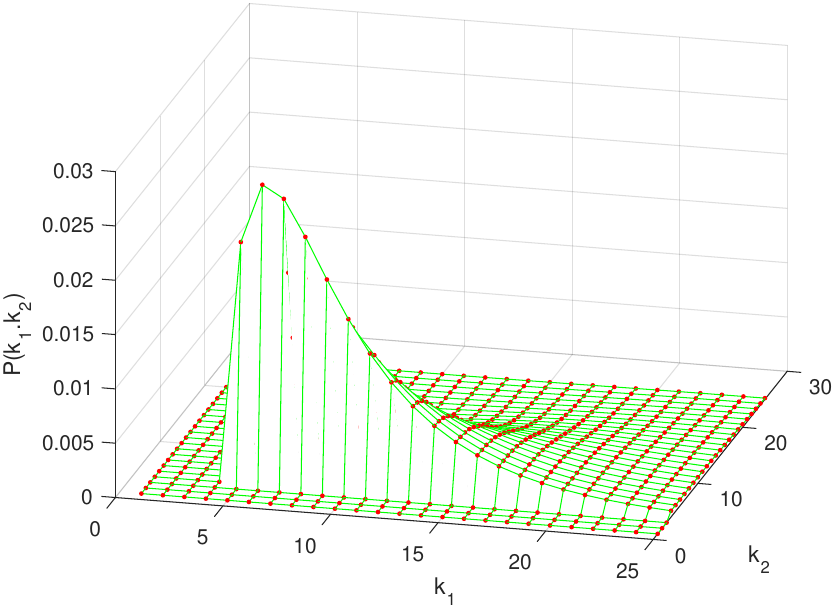}
\end{minipage}
\caption{degree correlation simulation results vs numerical calculation results.}
\label{fig:estimation}
\end{figure}

\section{undirected networks}

\subsection{state transition equation of undirected networks}
For undirected networks, we can obtain the final state transition equation for degree correlation $P_{[L,(k_1,k_2)]}(t+1)$ by merging the directions of edges as follows.\\
$(1)$ For $k_1=m, k_2=m$:
\begin{equation}
\begin{aligned}
P_{[L,(m,m)]}(t+1)&=P_{[L-m,(m-1,m)]}(t)\times P_{[L-m,(m-1,m)][L,(m,m)]}(t)\\
&\indent+P_{[L-m,(m-1,m-1)]}(t)\times P_{[L-m,(m-1,m-1)][L,(m,m)]}(t) \\
&\indent+P_{[L-m,(m,m)]}(t)\times P_{[L-m,(m,m)][L,(m,m)]}(t)\\
&\indent+\sum_{k_i,k_j}P_{[L-m,(k_i,k_j)]}(t)\times P_{[L-m,(k_i,k_j)][L,(m,m)]}(t);
\end{aligned}
\end{equation}
and
\begin{equation}
\begin{aligned}
P_{[L,(m,m)]}(t+1)&=P_{[L-m,(m,m-1)]}(t)\times P_{[L-m,(m,m-1)][L,(m,m)]}(t)\\
&\indent+P_{[L-m,(m-1,m-1)]}(t)\times P_{[L-m,(m-1,m-1)][L,(m,m)]}(t) \\
&\indent+P_{[L-m,(m,m)]}(t)\times P_{[L-m,(m,m)][L,(m,m)]}(t)\\
&\indent+\sum_{k_i,k_j}P_{[L-m,(k_j,k_i)]}(t)\times P_{[L-m,(k_j,k_i)][L,(m,m)]}(t).
\end{aligned}
\end{equation}
Hence, 

\begin{equation}
\begin{aligned}
P_{\overline{[L,(m,m)]}}(t+1)&=P_{\overline{[L-m,(m,m-1)]}}(t)\times P_{[L-m,(m,m-1)][L,(m,m)]}(t)\\
&\indent+P_{\overline{[L-m,(m-1,m-1)]}}(t)\times P_{[L-m,(m-1,m-1)][L,(m,m)]}(t) \\
&\indent+P_{\overline{[L-m,(m,m)]}}(t)\times P_{[L-m,(m,m)][L,(m,m)]}(t)\\
&\indent+\sum_{k_i,k_j}P_{\overline{[L-m,(k_j,k_i)]}}(t)\times P_{[L-m,(k_j,k_i)][L,(m,m)]}(t)\\
&=\frac{L-m}{L}\frac{{N-2\choose m-1}}{{N\choose m}} P_{\overline{[L-m,(m,m-1)]}}(t)+\frac{L-m}{L}\frac{{N-2\choose m-2}}{{N\choose m}}P_{\overline{[L-m,(m-1,m-1)]}}(t)\\
&\indent+\frac{L-m}{L}\frac{{N-2\choose m}}{{N\choose m}}P_{\overline{[L-m,(m,m)]}}(t)+\frac{L-m}{L}\frac{N_{m-1}}{N}.
\end{aligned}
\end{equation}

Similarly, we determine the transition equation of state $P_{[L,(k_1,k_2)]}(t+1)$ for other value of $k_1$ and $k_2$.\\
$(2)$ For $k_1=m, k_2=m+1$:
\begin{equation}
\begin{aligned}
 P_{\overline{[L,(m,m+1)]}}(t+1)&=P_{\overline{[L-m,(m-1,m+1)]}}(t)\times P_{[L-m,(m-1,m+1)][L,(m,m+1)]}(t)\\
 &\indent +P_{\overline{[L-m,(m,m)]}}(t)\times P_{[L-m,(m,m)][L,(m,m+1)]}(t)\\
 &\indent +P_{\overline{[L-m,(m-1,m)]}}(t)\times P_{[L-m,(m-1,m)][L,(m,m+1)]}(t)\\
 &\indent +P_{\overline{[L-m,(m,m+1)]}}(t)\times P_{[L-m,(m,m+1)][L,(m,m+1)]}(t)\\
 &\indent +\sum_{k_i,k_j}P_{\overline{[L-m,(k_i,k_j)]}}(t)\times P_{[L-m,(k_i,k_j)][L,(m,m+1)]}(t)\\
 &=\frac{L-m}{L}\frac{{N-2\choose m-1}}{{N\choose m}}P_{\overline{[L-m,(m-1,m+1)]}}(t)+\frac{L-m}{L}\frac{{N-2\choose m-1}}{{N\choose m}}P_{\overline{[L-m,(m,m)]}}(t)\\
 &\indent +\frac{L-m}{L}\frac{{N-2\choose m-2}}{{N\choose m}}P_{\overline{[L-m,(m-1,m)]}}(t)+\frac{L-m}{L}\frac{{N-2\choose m}}{{N\choose m}}P_{\overline{[L-m,(m,m+1)]}}(t)\\
 &\indent+\frac{m}{L}\frac{N_m}{N}.  
\end{aligned}
\end{equation}
$(3)$ For $k_1=m,k_2>m+1$:
\begin{equation}
\begin{aligned}
P_{\overline{[L,(m,k_2)]}}(t+1)&=P_{\overline{[L-m,(m-1,k_2)]}}(t)\times P_{[L-m,(m-1,k_2)][L,(m,k_2)]}(t)\\
&\indent +P_{\overline{[L-m,(m,k_2-1)]}}(t)\times P_{[L-m,(m,k_2-1)][m,k_2]}(t)\\
&\indent +P_{\overline{[L-m,(m-1,k_2-1)]}}(t)\times P_{[L-m,(m-1,k_2-1)][L,(m,k_2)]}(t)\\
&\indent +P_{\overline{[L-m,(m,k_2)]}}(t)\times P_{[L-m,(m,k_2)][L,(m,k_2)]}(t)\\
&\indent +\sum_{k_i,k_j}P_{\overline{[L-m,(k_i,k_j)]}}(t)\times P_{[L-m,(k_i,k_j)][L,(m,k_2)]}(t)\\
&=\frac{L-m}{L}\frac{{N-2\choose m-1}}{{N\choose m}}P_{\overline{[L-m,(m-1,k_2)]}}(t)+\frac{L-m}{L}\frac{{N-2\choose m-1}}{{N\choose m}}P_{\overline{[L-m,(m,k_2-1)]}}(t)\\
&\indent +\frac{L-m}{L}\frac{{N-2\choose m-2}}{{N\choose m}}P_{\overline{[L-m,(m-1,k_2-1)]}}(t)+\frac{L-m}{L}\frac{{N-2\choose m}}{{N\choose m}}P_{\overline{[L-m,(m,k_2)]}}(t)\\
 &\indent+\frac{m}{L}\frac{N_{k_2-1}}{N}.  
\end{aligned}
\end{equation}

$(4)$ For $k_1>m, k_2=k_1$:
\begin{equation}
\begin{aligned}
P_{\overline{[L,(k_1,k_2)]}}(t+1)&=P_{\overline{[L-m,(k_1-1,k_2)]}}(t)\times P_{[L-m,(k_1-1,k_2)][L,(k_1,k_2)]}(t)\\
&\indent +P_{\overline{[L-m,(k_2-1,k_1)]}}(t)\times P_{[L-m,(k_2-1,k_1)][L,(k_2,k_1)]}(t)\\
&\indent +P_{\overline{[L-m,(k_1-1,k_2-1)]}}(t)\times P_{[L-m,(k_1-1,k_2-1)][L,(k_1,k_2)]}(t)\\
&\indent +P_{\overline{[L-m,(k_1,k_2)]}}(t)\times P_{[L-m,(k_1,k_2)][L,(k_1,k_2)]}(t)\\
&=\frac{L-m}{L}\frac{{N-2\choose m-1}}{{N\choose m}}P_{\overline{[L-m,(k_1-1,k_2)]}}(t)+\frac{L-m}{L}\frac{{N-2\choose m-1}}{{N\choose m}}P_{\overline{[L-m,(k_2-1,k_1)]}}(t)\\
&\indent +\frac{L-m}{L}\frac{{N-2\choose m-2}}{{N\choose m}}P_{\overline{[L-m,(k_1-1,k_2-1)]}}(t)+\frac{L-m}{L}\frac{{N-2\choose m}}{{N\choose m}}P_{\overline{[L-m,(k_1,k_2)]}}(t)
\end{aligned}
\end{equation}

$(5)$ For $k_1>m, k_2-k_1=1$:
\begin{equation}
\begin{aligned}
P_{\overline{[L,(k_1,k_2)]}}(t+1)&=P_{\overline{[L-m,(k_1-1,k_2)]}}(t)\times P_{[L-m,(k_1-1,k_2)][L,(k_1,k_2)]}(t)\\
&\indent +2P_{\overline{[L-m,(k_1,k_2-1)]}}(t)\times P_{[L-m,(k_1,k_2-1)][L,(k_1,k_2)]}(t)\\
&\indent +P_{\overline{[L-m,(k_1-1,k_2-1)]}}(t)\times P_{[L-m,(k_1-1,k_2-1)][L,(k_1,k_2)]}(t)\\
&\indent +P_{\overline{[L-m,(k_1,k_2)]}}(t)\times P_{[L-m,(k_1,k_2)][L,(k_1,k_2)]}(t)\\
&= \frac{L-m}{L}\frac{{N-2\choose m-1}}{{N\choose m}}P_{\overline{[L-m,(k_1-1,k_2)]}}(t)+2\frac{L-m}{L}\frac{{N-2\choose m-1}}{{N\choose m}}P_{\overline{[L-m,(k_1,k_2-1)]}}(t)\\
&\indent+\frac{L-m}{L}\frac{{N-2\choose m-2}}{{N\choose m}}P_{\overline{[L-m,(k_1-1,k_2-1)]}}(t)+\frac{L-m}{L}\frac{{N-2\choose m}}{{N\choose m}}P_{\overline{[L-m,(k_1,k_2)]}}(t)
\end{aligned}
\end{equation}

$(6)$ For $k_1>m, k_2-k_1>1$:
\begin{equation}
\begin{aligned}
P_{\overline{[L,(k_1,k_2)]}}(t+1)&=P_{\overline{[L-m,(k_1-1,k_2)]}}(t)\times P_{[L-m,(k_1-1,k_2)][L,(k_1,k_2)]}(t)\\
&\indent +P_{\overline{[L-m,(k_1,k_2-1)]}}(t)\times P_{[L-m,(k_1,k_2-1)][L,(k_1,k_2)]}(t)\\
&\indent +P_{\overline{[L-m,(k_1-1,k_2-1)]}}(t)\times P_{[L-m,(k_1-1,k_2-1)][L,(k_1,k_2)]}(t)\\
&\indent +P_{\overline{[L-m,(k_1,k_2)]}}(t)\times P_{[L-m,(k_1,k_2)][L,(k_1,k_2)]}(t)\\
&=\frac{L-m}{L}\frac{{N-2\choose m-1}}{{N\choose m}}P_{\overline{[L-m,(k_1-1,k_2)]}}(t)+\frac{L-m}{L}\frac{{N-2\choose m-1}}{{N\choose m}}P_{\overline{[L-m,(k_1,k_2-1)]}}(t)\\
&\indent +\frac{L-m}{L}\frac{{N-2\choose m-2}}{{N\choose m}}P_{\overline{[L-m,(k_1-1,k_2-1)]}}(t)+\frac{L-m}{L}\frac{{N-2\choose m}}{{N\choose m}}P_{\overline{[L-m,(k_1,k_2)]}}(t)
\end{aligned}
\end{equation}

\subsection{the steady degree correlation equation}
After taking the limit operation, we obtained the stable correlation results of the undirected network as follows:

\begin{equation}
P\overline{(k_1,k_2)}=\left\{
\begin{aligned}
&0,\quad\quad(k_1=k_2=m)\\
&\frac{m}{2m+1}P\overline{(m,k_2-1)}+\frac{1}{2m+1}P(k_2-1),\quad\quad(k_1=m,k_2\geq k_1+1)\\
&\frac{m}{2m+1}P\overline{(k_1-1,k_2)},\quad\quad(k_1>m,k_2=k_1)\\
&\frac{m}{2m+1}P\overline{(k_1-1,k_2)}+\frac{2m}{2m+1}P\overline{(k_1,k_2-1)},\quad\quad(k_1>m,k_2-k_1=1)\\
&\frac{m}{2m+1}P\overline{(k_1-1,k_2)}+\frac{m}{2m+1}P\overline{(k_1,k_2-1)}.\quad\quad (k_1>m,k_2=k_1+1)\\
\end{aligned}
\right.
\end{equation}

Referring to the degree correlation determination method based on the generating function in the directed network part, here we directly give the recursive relationship between the generating functions in the undirected network, and the final degree correlation is the coefficient of the corresponding function.\\
\begin{equation}
\left\{
\begin{aligned}
&G_{m}(x)=\frac{1}{m+1}\sum_{k=m+1}^{\infty}{\sum_{i=1}^{k-m}\frac{m^{k-m-1}}{(2m+1)^i(m+1)^{k-m-i}}}x^k\\
&G_{m+1}(x)=\frac{m}{2m+1-mx}G_m(x)+P\overline{(m,m+1)}x^m+\frac{m}{2m+1-mx}P\overline{(m+1,m+1)}x^{m+2}\\
&G_{m+2}(x)=\frac{m}{2m+1-mx}G_{m+1}(x)+[P\overline{(m,m+2)}-\frac{m}{2m+1-mx}P\overline{(m,m+1)}]x^m\\
&\indent\indent +[P\overline{(m+1,m+2)}-\frac{m}{2m+1-mx}P\overline{(m+1,m+1)}]x^{m+1}+\frac{m}{2m+1-mx}P\overline{(m+2,m+2)}x^{m+3}\\
&...
\end{aligned}
\right.
\end{equation}

\subsection{numerical simulation}

\begin{table}                                                                                
\centering                                                                                   
\begin{tabular}{|c|c|c|c|c|c|c|}                                                               
\hline      
$(k_1,k_2)$ & 1 & 2 & 3 & 4 & 5 & 6\\
\hline 
1 & 0 & 5.66e-05 & 1.01e-04 & 2.12e-04 & 1.14e-04 & 1.63e-04 \\           
\hline                                                                                       
2 & 5.66e-05 & 6.22e-05 & 7.53e-05 & 1.18e-04 & 2.32e-05 & 8.37e-05 \\
\hline                                                                                       
3 & 1.01e-04 & 7.53e-05 & 1.67e-04 & 1.20e-04 & 2.13e-04 & 9.57e-06 \\ 
\hline                                                                                       
4 & 2.12e-04 & 1.18e-04 & 1.20e-04 & 3.31e-05 & 1.38e-04 & 6.33e-06 \\ 
\hline                                                                                       
5 & 1.14e-04 & 2.32e-05 & 2.13e-04 & 1.38e-04 & 1.13e-04 & 1.34e-04 \\ 
\hline                                                                                       
6 & 1.63e-04 & 8.38e-05 & 9.57e-06 & 6.33e-06 & 1.34e-04 & 3.57e-05 \\  
\hline                                                                                       
\end{tabular}                                                                                
\caption{Numerical calculation error for m=1.}              
\label{table:MyTableLabel}                                
\end{table}

\begin{table}                                              
\centering                                                 
\begin{tabular}{|c|c|c|c|c|c|c|}                             
\hline     
$(k_1,k_2)$ & 2 & 3 & 4 & 5 & 6 & 7\\
\hline 
2 & 0 & 3.61e-05 & 1.99e-04 & 1.04e-04 & 3.20e-05 & 2.04e-04 \\   \hline                                                     
3 & 3.62e-05 & 4.08e-05 & 8.42e-05 & 1.70e-04 & 1.50e-04 & 9.21e-05 \\ 
\hline                                                     
4 & 1.99e-04 & 8.42e-05 & 2.53e-04 & 1.56e-04 & 1.15e-05 & 2.79e-05 \\    
\hline                                                     
5 & 1.04e-04 & 1.70e-04 & 1.56e-04 & 2.33e-04 & 1.44e-04 & 2.14e-05 \\   
\hline                                                     
6 & 3.20e-04 & 1.50e-04 & 1.15e-04 & 1.44e-04 & 8.72e-05& 2.89e-04 \\ 
\hline                                                     
7 & 2.04e-04 & 9.20e-05 & 2.79e-05 & 2.14e-05  & 2.88e-04 & 1.30e-04 \\
\hline                                                     
\end{tabular}                                              
\caption{Numerical calculation error for m=2.}                                   
\label{table:MyTableLabel}                                 
\end{table} 
\begin{table}                                               
\centering                                                  
\begin{tabular}{|c|c|c|c|c|c|c|}                              
\hline    
$(k_1,k_2)$ & 3 & 4 & 5 & 6 & 7 & 8\\
\hline 
3 & 0 & 1.39e-04 & 8.24e-05 & 6.44e-05 & 1.13e-04 & 1.20e-04 \\   \hline                                                      
4 & 1.39e-04 & 7.18e-05 & 2.01e-05 & 2.38e-04 & 1.23e-05 & 8.09e-05 \\   
\hline                                                      
5 & 8.24e-05 & 2.01e-05 & 3.83e-05 & 1.15e-05 & 1.48e-04 & 1.46e-04 \\
\hline                                                      
6 & 6.44e-05 & 2.38e-04 & 1.15e-05 & 1.23e-04 & 1.98e-05 & 1.40e-04 \\  
\hline                                                      
7 & 1.13e-04 & 1.23e-05 & 1.48e-04 & 1.98e-05 & 4.12e-05 & 1.16e-04 \\   
\hline                                                      
8 & 1.20e-04 & 8.08e-05 & 1.46e-04 & 1.40e-04 & 1.16e-04 & 1.05e-04 \\
\hline                                                      
\end{tabular}                                               
\caption{Numerical calculation error for m=3.}                                    
\label{table:MyTableLabel}                                  
\end{table}      

\begin{table}                                               
\centering                                                  
\begin{tabular}{|c|c|c|c|c|c|c|}                              
\hline  
$(k_1,k_2)$ & 4 & 5 & 6 & 7 & 8 & 9\\
\hline 
4 & 0 & 3.11e-05 & 7.43e-05 & 9.76e-05 & 3.16e-05 & 5.70e-05 \\   \hline                                                      
5 & 3.11e-05 & 2.88e-05 & 1.51e-04 & 8.18e-05 & 2.27e-05 & 1.09e-04 \\   
\hline                                                      
6 & 7.43e-05 & 1.51e-04 & 7.17e-05 & 8.49e-05 & 3.21e-05 & 8.21e-05\\    
\hline                                                      
7 & 9.76e-05 & 8.18e-05 & 8.49e-05 & 1.70e-05 & 1.33e-04& 8.22e-05 \\
\hline                                                      
8 & 3.16e-05 & 2.27e-05 & 3.21e-05 & 1.33e-04 & 3.28e-06 & 7.10e-05 \\  
\hline                                                      
9 & 5.70e-05 & 1.09e-04 & 8.21e-05 & 8.22e-05 & 7.11e-05 & 3.61e-05\\ 
\hline                                                      
\end{tabular}                                               
\caption{Numerical calculation error for m=4.}                                    
\label{table:MyTableLabel}                                  
\end{table}

\begin{figure}[htbp]
	\centering     
 \begin{minipage}{0.49\linewidth}
 \centering
 \includegraphics[width=0.9\linewidth]{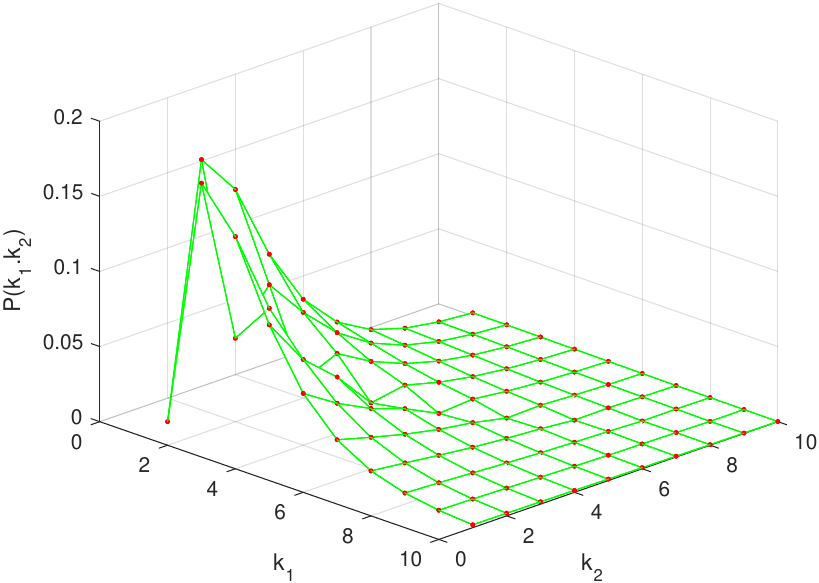}
 \end{minipage}
\begin{minipage}{0.49\linewidth}
 \centering
 \includegraphics[width=0.9\linewidth]{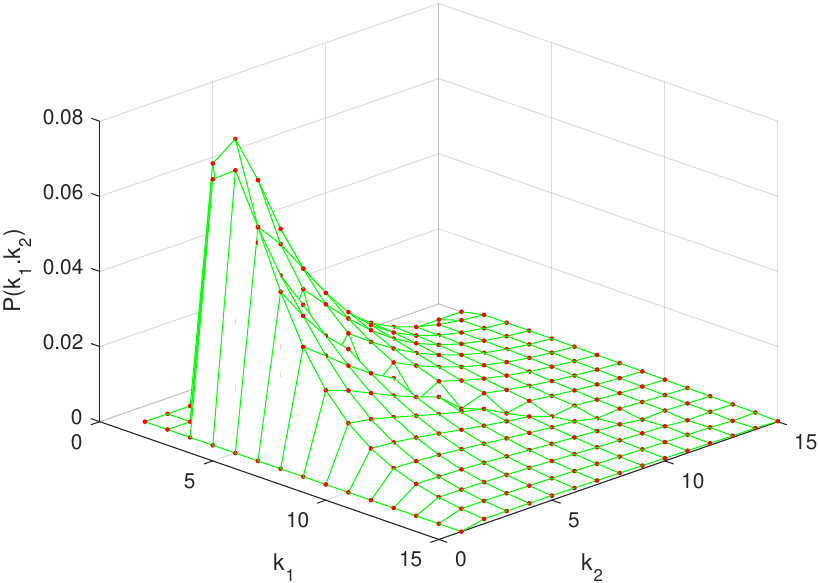}
\end{minipage}
\qquad
 \begin{minipage}{0.49\linewidth}
 \centering
 \includegraphics[width=0.9\linewidth]{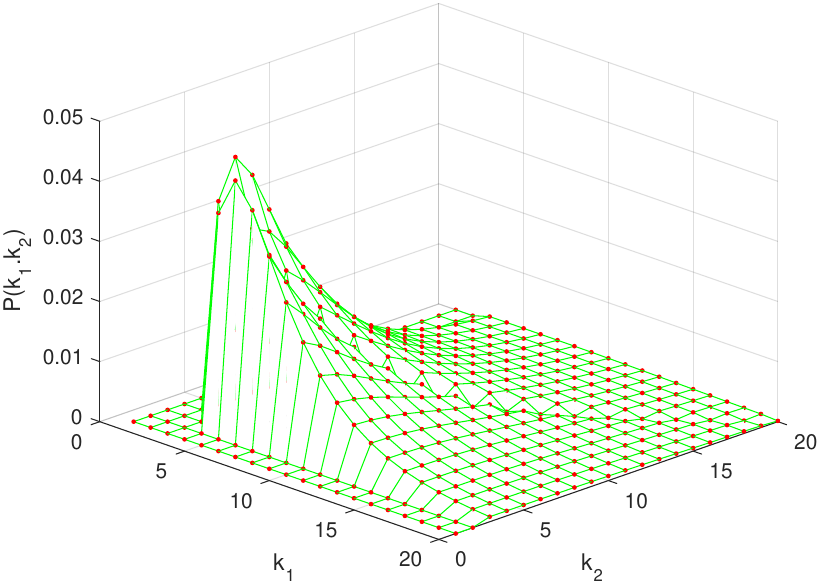}
\end{minipage}
 \begin{minipage}{0.49\linewidth}
 \centering
 \includegraphics[width=0.9\linewidth]{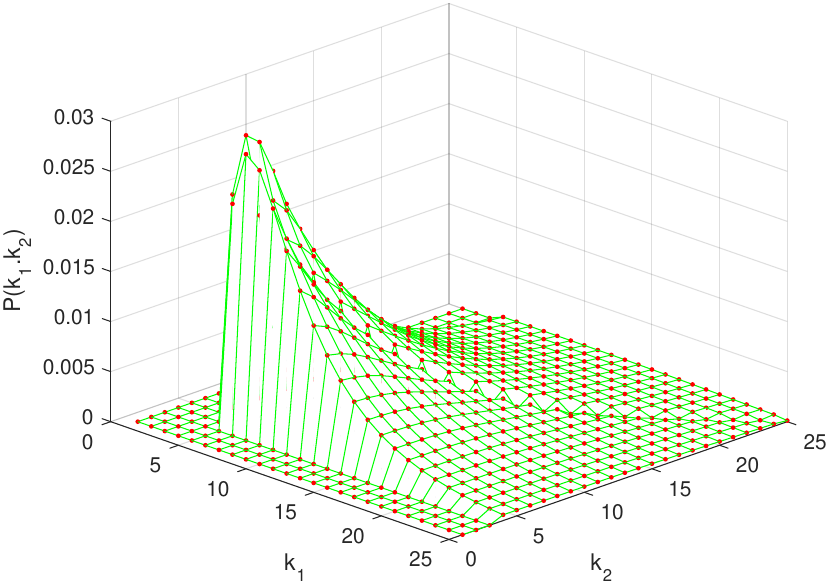}
\end{minipage}
\caption{degree correlation simulation results vs numerical calculation results.}
\label{fig:estimation}
\end{figure}

According to the calculation errors in the table, the difference between the degree correlation obtained from 100 simulated networks and the numerical calculation results obtained from the recursive relationship is very small, indicating the correctness of our modeling method and the derivation results. Therefore, it is feasible to derive the degree correlation directly from the recursive relationship of the degree correlation of undirected networks. It can be seen from the three-dimensional mesh of degree correlation presented in the figure that under the setting of the evolution mechanism of the pure growth exponential network, the corresponding model degree correlation for different parameters $m$ still shows some differences. Specifically, when the degree correlation of the evolving network tends to be stable, its non-zero degree correlation starts from $P(m,m)$, that is, the degree correlation peak value directly associated with the $m$ value gradually shifts away from the origin of the coordinate as $m$ increases. It can also be seen from the figure that the degree correlation distribution of the undirected network is symmetric about the diagonal element $P(k_1,k_2), k_1=k_2$, which is different from the single-peak state presented by the degree correlation distribution of the directed network.

\section{conclusion} 
Degree correlation, as an important statistic that connects network topology with many network dynamic characteristics based on structure, has always been a research focus in the field of network science. However, due to the lack of practical research methods, the evolutionary characteristics of degree correlation and the accompanying dynamic influence of evolving networks that are more widely present in reality have never been fully explored and applied. Based on this situation, this paper proposes a modeling method based on the Markov chain for the degree correlation of evolving networks under pure growth mechanism. It gives theoretical results of degree correlation for directed and undirected networks, respectively. The correctness of the modeling method and recursive results adopted in this paper are verified by comparing the simulation experiments with the numerical calculation results, and the characteristics of the network degree correlation distribution under directed and undirected conditions are analyzed and explained, respectively. \\
\indent The research in this paper is the first step in the theoretical study of the degree correlation of evolving networks. In the future, we can also discuss degree correlation under more complex evolutionary mechanisms based on this paper, such as the study of degree correlation under the power law distribution generation mechanism, and introduce parameters regulation mechanism to achieve network control. At the same time, regarding the mutual influence between degree distribution and degree correlation in the evolution process, when the degree correlation change is solved, the degree distribution analysis and solution of evolving networks containing node deletion can also be solved. In addition to the above-mentioned complex network forward problems that can be studied in depth, this work is also expected to provide theoretical support for subsequent inverse problems in this field, such as reasoning about evolutionary mechanisms from network topological structures.

\section*{Acknowledgment}
This work was supported by the National Natural Science Foundation of China (NO. 61273015; NO. 62371094).\\

\bibliographystyle{plain}
\bibliography{reference}

\ifCLASSOPTIONcaptionsoff
  \newpage
\fi

\end{document}